\documentclass[conference,letterpaper]{IEEEtran}

\def\BibTeX{{\rm B\kern-.05em{\sc i\kern-.025em b}\kern-.08em
		T\kern-.1667em\lower.7ex\hbox{E}\kern-.125emX}}

\usepackage{microtype}
\usepackage{graphicx}
\usepackage{subcaption}
\usepackage{booktabs} 

\usepackage[utf8]{inputenc} 
\usepackage[T1]{fontenc}
\usepackage{url}
\usepackage{ifthen}
\usepackage[noadjust]{cite}
\usepackage[cmex10]{amsmath} 
\usepackage{amsthm,amssymb,amsfonts}
\usepackage{algorithm}
\usepackage[noend]{algpseudocode}
\usepackage{graphicx}
\usepackage{textcomp}
\usepackage{xcolor}
\usepackage{comment}
\usepackage{epstopdf}
\usepackage{float}
\usepackage{amssymb}
\usepackage{color}
\usepackage{relsize}
\usepackage{enumitem}
\usepackage{cleveref}
\usepackage{cases}
\usepackage[nocomma]{optidef}
\usepackage[mode=buildnew]{standalone}
\usepackage{breqn}
\usepackage{diagbox}

\usepackage{caption}
\setlist{leftmargin=4.1mm}

\theoremstyle{plain}
\newtheorem{theorem}{Theorem}
\newtheorem{lemma}{Lemma}

\newtheorem{remark}{Remark}

\theoremstyle{definition}

\newcommand{\bfA}{\mathbf{A}}
\newcommand{\bfB}{\mathbf{B}}

\newcommand{\bfq}{\mathbf{q}}
\newcommand{\bfU}{\mathbf{U}}
\newcommand{\bfC}{\mathbf{C}}
\newcommand{\bfOMG}{\mathbf{\Omega}}
\newcommand{\bfz}{\mathbf{Z}}

\newcommand{\ds}{\mathbf{C}_{\mathcal{S}}}
\newcommand{\dmn}{\mathbf{C}^{mn}_{\mathcal{S}}}

\newcommand{\cq}{\mathcal{Q}}

\newcommand{\calS}{\mathcal{S}}
\newcommand{\calA}{\mathcal{A}}
\newcommand{\calB}{\mathcal{B}}
\newcommand{\calT}{\mathcal{T}}

\newcommand{\tA}{\widetilde{\bfA}}
\newcommand{\tB}{\widetilde{\bfB}}
\newcommand{\tS}{\widetilde{\calS}}

\newcommand{\hA}{\widehat{\bfA}}
\newcommand{\hB}{\widehat{\bfB}}

\DeclarePairedDelimiter\floor{\lfloor}{\rfloor}

\DeclareSymbolFont{mathptmxlargesymbols}{OMX}{ztmcm}{m}{n}
\DeclareMathSymbol{\upsumop}{\mathop}{mathptmxlargesymbols}{"50}

\newcommand\lev[1]{{\color{black}#1}}
\newcommand\levi[1]{{\color{black}#1}}
\newcommand\levj[1]{{\color{black}#1}}

\begin{document}
	
\title{Fully Private Grouped Matrix Multiplication \\
	 with Colluding Workers
	 \vspace{-0.5cm}
}

\author{\IEEEauthorblockN{Lev Tauz, Lara Dolecek}
\IEEEauthorblockA{Department of Electrical and Computer Engineering, University of California, Los Angeles\\
	levtauz@ucla.edu, dolecek@ee.ucla.edu 
	\vspace{-0.75cm}
}
}

\maketitle

\begin{abstract}
In this paper, we present a novel variation of the coded matrix multiplication problem which we refer to as fully private grouped matrix multiplication (FPGMM). In FPGMM, a master wants to compute a group of matrix products between two matrix libraries that can be accessed by all workers \levj{while ensuring that any number of prescribed colluding workers learn nothing about which matrix products the master desires, nor the number of matrix products}. We present an achievable scheme using a variant of Cross-Subspace Alignment (CSA) codes that offers flexibility in communication and computation cost. Additionally, we demonstrate how our scheme can outperform naive applications of schemes used in a related privacy focused coded matrix multiplication problem.
\end{abstract}
\vspace{-0.15cm}
\section{Introduction and Motivation}
\vspace{-0.15cm}

Matrix multiplication is a major building block of many modern big data applications such as machine learning or data analysis. With the rise of Big Data, matrices have gotten so large that their multiplication must be done on a distributed system of many workers. Unfortunately, outsourcing the work across workers comes with additional concerns such as the presence of \textit{stragglers}  (i.e., workers that fail or are slow to respond) \cite{dean2013Tail}, hampering the speed of the system or the privacy concerns about the data. Coded computation is a field of research that tackles these issues utilizing techniques from channel coding for a variety of system models \cite{kim2019private,chang2019upload,chang2021capacity,jia2020x,aliasgari2020private,yang2021private,zhu2022systematic,zhu2022private,kim2020fully,hong2023straggler,yu2020entangled,zhu2021improved,zhu2020secure,chen2021gcsa}. For example, \textit{secure and private matrix multiplication} (SPMM) \cite{kim2019private,chang2019upload,chang2021capacity,jia2020x,aliasgari2020private,yang2021private} tasks a system to compute the product of a private matrix $\bfA$ with a specific matrix $\bfB_j$, $1 \leq j \leq k,$ among a library of matrices $\{\bfB_1,\dots,\bfB_k\}$ which are stored at the workers. To ensure privacy, the workers must not be able to learn anything about the matrix $\bfA$ and the index $j$ while collectively computing $\bfA\bfB_{j}$. Such a problem statement is highly reminiscent of the problem of \textit{private information retrieval} (PIR) \cite{chor1998private} where a master wants to extract data from a set of servers without revealing what data the master requires.  Expanding further upon the ideas of PIR, many variations of this problem have been proposed in literature. One example is  \textit{fully private matrix multiplication} (FPMM) \cite{zhu2022systematic,zhu2022private,kim2020fully,hong2023straggler,yu2020entangled} where the workers store two libraries of matrices and the master \levj{tasks the workers to privately calculate the product of two desired matrices from the shared libraries while being oblivious to the indices of the desired matrices}. Another variation is \textit{secure batch matrix multiplication} (SBMM) \cite{yu2020entangled,zhu2021improved,zhu2020secure,chen2021gcsa} where the master tasks the system to calculate the product of multiple matrix pairs without the workers learning anything about the matrices. In SBMM, the master stores and encodes the data that it sent to the workers that compute the desired matrix products without learning anything about the input matrices.

In this work, we propose a brand new variation that is a generalization of FPMM, which we refer to as \textit{fully private grouped matrix multiplication} (FPGMM) where a master can request multiple matrix products, i.e., a group of products, in a single request and the master wishes to preserve the privacy of their request. This new problem can also be seen as a variation of SBMM by allowing for a different privacy constraint on batch matrix multiplication, though we go a little further by requiring the batch size to be private as well. One can imagine many practical scenarios where a master may wish to request multiple matrix products. For example, consider the scenario of a recommender system based on collaborative filtering where recommendations are created by computing the product of two matrices, one describing the profile of the user and one representing the profile of the items to recommend. The master may wish to calculate the product recommendations for a variety of users and items without revealing to the system which users and items were considered. Now, one may think that this problem can be solved by sequentially applying a solution for the simpler FPMM problem and get each individual matrix product one at the time over multiple rounds. \levj{Yet, requesting multiple distinct matrix products reveals to the workers how many products are desired, reducing the privacy of the request. For example, if there are 10 matrix products and the master requests 5, then the workers know that each matrix could have been requested with probability $\frac{1}{2}$}. In the extreme case where the master wants to calculate all pair-wise matrix products, the workers know with certainty what was requested by the master. To ensure that this information is not leaked using the multiple-round FPMM scheme, the system must provide an extra layer of anonymity so that workers cannot associate a group of computations to a single user, which may add significant overhead. Thus, there is significant merit in studying FPGMM and creating a privacy preserving system without the additional anonymity overhead. 


To solve the problem of FPGMM, we present a new achievable scheme based on the idea of Cross-Subspace Alignment (CSA) codes \cite{jia2020x,jia2021cross,chen2021gcsa} which utilize rational functions to encode the data. \lev{CSA codes have already been utilized to solve the problem of FPMM \cite{kim2020fully} and SBMM \cite{zhu2021improved,chen2021gcsa}. We will demonstrate a new scheme that allows for flexibility in the communication cost, computation cost, and straggler resilience while allowing for information-theoretic privacy from up to a fixed amount of colluding workers. Additionally, we will demonstrate that straightforward application of CSA codes designed for other variants cannot be applied to FPGMM due to the new privacy considerations. }

The paper is organized as follows. We present the system model in Section \ref{sec:prelim}. \lev{We demonstrate an illustrative examples in Section \ref{sec:example_disc} and discuss the major points of our scheme.}  In Section \ref{sec:main_res}, we state our main result and discuss its implications. We provide our novel scheme in Section \ref{sec:scheme}. Finally, we provide concluding remarks in Section \ref{sec:conclusion}.

\textit{Notation:} We denote an integer set from $1$ to $N$ as $[N]$. 
Given a subset $S\subseteq [N]$, we define $x_{S} \triangleq \{x_i: i \in S\}$. 
Given two sets $A$ and $B$, $A \times B$ is the Cartesian product of the two sets. $H(X)$ and $I(X;Y)$ denote the standard information entropy and mutual information in terms of $q$-ary units.
\vspace{-0.15cm}
\section{System Model and Preliminaries}\label{sec:prelim}
\vspace{-0.15cm}
\lev{
We now introduce the problem setting for FPGMM. Assume a distributed system with one master and $N$ workers. All workers store two libraries of matrices $\bfA_{[L_A]} = \{\bfA_{i} \in \mathbb{F}_q^{\alpha \times \alpha }, \forall i \in [L_A]\}$ and $\bfB_{[L_B]} = \{\bfB_{i} \in \mathbb{F}_q^{\alpha \times \alpha},  \forall i \in [L_B]\}$ where  $\mathbb{F}_q$ is a finite field of size $q$.\footnote{We note that our schemes are applicable for non-square matrices and that we focus on square matrices only for notational convenience.} We assume that the matrices in the libraries are statistically independent (see \cite{kim2020fully}). 
Given a set $\mathcal{S} \subseteq [L_A] \times [L_B]$, the master wants to obtain the matrix products $\ds \triangleq \{\mathbf{A}_i\mathbf{B}_j: (i,j) \in \mathcal{S}\}$. We assume that $\mathcal{S}$ is equally likely to be any non-empty subset of $[L_A] \times [L_B]$ and is chosen independently of the data stored in the two matrix libraries, i.e., $I(\mathcal{S}; \bfA_{[L_A]},\bfB_{[L_B]}) = 0$. The master does not want the workers to learn anything about $\calS$.

The FPGMM scheme contains the following phases:
\begin{itemize}
	\item Encoding Phase: The master designs queries $\bfq_{i}, i \in [N]$  based on $\calS$.
	\item Query and Computation: The master sends query  $\bfq_{i}, i \in [N]$ to worker $i$. Worker $i$ then uses  $\bfq_{i}$ to encode the data libraries using a function $f(\bfA_{[L_A]},\bfB_{[L_B]},\bfq_i)=\bfU_i$ and outputs $\bfU_i$.
	\item Reconstruction: The master downloads $\bfU_i$ from the servers. Some workers may be stragglers and fail to respond. The master attempts to reconstruct $\ds$ from the responding servers. 
\end{itemize}
Additionally, FPGMM requires that $\mathcal{S}$ is kept private from up to $T$ colluding workers. This privacy requirement includes both the  elements and cardinality of $\calS$. Formally, FPGMM requires
\vspace{-0.1cm}
\begin{equation}\label{eq:privacy_req}
	I(\mathcal{S}; \mathbf{q}_{\mathcal{T}}, \bfA_{[L_A]}, \bfB_{[L_B]}) = 0, \forall \mathcal{T} \in [N], |\mathcal{T}| \leq T.
\end{equation}
The described model is summarized in Fig. \ref{fig:model}.

\begin{figure}
	\centering
	\includegraphics[page=4, width=0.70\linewidth]{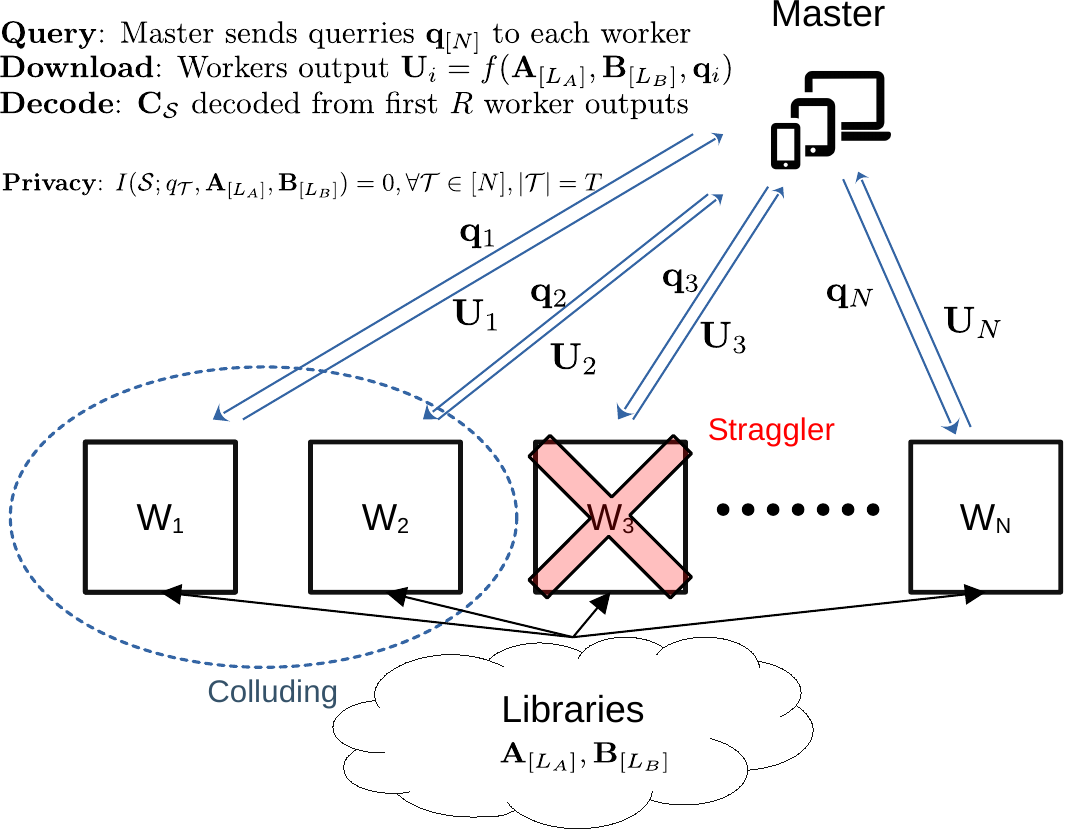}
	\caption{System model of FPGMM.}\label{fig:model}
	\vspace{-0.5cm}
\end{figure}

For FPGMM, the three important performance metrics are the following:
\begin{itemize}
	\item \textit{Recovery Threshold R}: The minimum number of worker outputs needed in order to reconstruct $\ds$. Specifically, given a response set $\mathcal{R} \subseteq [N]$, $\ds$ can be reconstructed from $\{\bfU_i\}_{i\in \mathcal{R}}$ if $|\mathcal{R}| \geq R$.
	\item \textit{Normalized Computational Complexity (NCC) C}: The average order of the number of arithmetic operations required to compute the function $f$ at each worker, normalized by $|\calS|\alpha^3$, which is the standard computational complexity of computing the $|\calS|$ matrix products.
	\item \textit{Normalized Download Cost (NCC) D}: The total number of symbols retrieved by the master normalized by the number of symbols in $\ds$. Formally, given a recovery threshold $R$,
	\begin{align}
		D = \max_{\mathcal{R} \in \binom{[N]}{R}}\frac{\sum_{i\in \mathcal{R}}|U_i|}{|\ds|} = \max_{\mathcal{R} \in \binom{[N]}{R}}\frac{\sum_{i\in \mathcal{R}}|U_i|}{|\mathcal{S}|\alpha^2}.
	\end{align}
\end{itemize}
We do not consider the upload cost of $\bfq_i$ in this work since we assume that the matrices are large, i.e. $\alpha \gg L_A,L_B$, and, thus, the upload cost of the queries is negligible. 
}

Finally, we wish to highlight a special case of FPGMM where there are no stragglers which is named the non-straggler scenario. In this case, all workers contribute their results to the master, thus the recovery threshold must be equal to the number of workers. Additionally, without any stragglers, the important performance metrics are the NCC and NDC. Thus, it is important to find a good trade-off between these two metrics.

We rely on the following lemma about rational function interpolation as a fundamental building block of our code construction:
\vspace{-0.2cm}
\begin{lemma}\label{lemma:cauchy_vandermonde}
	(\cite{Gasca1989Poles}) Let $f_{1}, f_{2}, \cdots, f_{M}, x_1,x_2,\cdots,x_N$ be $M+N$ distinct elements of $\mathbb{F}_q$, with $|\mathbb{F}_q| \geq M+N$. Let $M+1 < N$. Then, the coefficients $e_{j}, j \in [M+N],$ of the following function can be interpolated from the function outputs of the $N$ evaluation points (i.e., $\{F(x_i):i\in[K]\}$):
	\vspace{-0.15cm}
	\begin{equation}
		F(z) = \sum_{i=1}^{M}\frac{e_{i}}{(z-f_i)} + \sum_{j=0}^{N-M-1}e_{j+M+1}z^j. \label{eq:cauchy_van_function}
	\end{equation}
\end{lemma}
\vspace{-0.35cm}
\section{Illustrative Example and Discussion} \label{sec:example_disc}
\vspace{-0.1cm}
Before presenting our proposed scheme, we wish to show an illustrative example to highlight the key components of our scheme. Assume that $L_A = L_B = 2$  and that $\mathcal{S} = \{(1,1),(1,2)\}$. As such, we want to retrieve $\ds = \{\bfA_1\bfB_1,\bfA_1\bfB_2\}$. Additionally, let $T=1$ to protect privacy against $1$ curious worker. The master specifies to the workers to partition the $\bfB_{[L_B]}$ data matrices as follows:
\begin{align*}
	\bfB_{j} = \begin{bmatrix}
		\bfB_{i,1} & \bfB_{i,2}
	\end{bmatrix}, \forall j \in [2].
\end{align*}
Now, to calculate $\{\bfA_1\bfB_1,\bfA_1\bfB_2\}$, a sufficient condition is to calculate $\{\bfA_{1}\bfB_{1,b}\}^{2}_{b=1}\cup \{\bfA_{1}\bfB_{2,b}\}^{2}_{b=1}$

Let $f_{1,1}, f_{1,2},f_{2,1},$ and $f_{2,2}$ be distinct elements from $F_q$. The master groups up the computations into two groups $\{\bfA_{1}\bfB_{1,1}, \bfA_{1}\bfB_{2,1}\}$ and $\{\bfA_{1}\bfB_{1,2}, \bfA_{1}\bfB_{2,2}\}$. Note that the grouping is arbitrary but the number of groups is carefully chosen. If the number of groups was instead $4$, then the workers can easily determine that $|\calS|=2$ due to knowledge of the partitioning parameters. This limits the straightforward applicability of CSA codes used for SBMM \cite{zhu2021improved,chen2021gcsa} for FPGMM because they are designed for grouping computations based on the size of the batches, i.e. dependent on $|\cal{S}|$. We address this issue by grouping computations based on the partitioning parameters which are chosen independent of $\calS$.

Consider the following encoding functions
\begin{align} 
	a_{i,k}(x) &= \omega_k(x)z^a_{i,k} + \notag \\
	&+ \omega_k(x) \times \begin{cases}
	\frac{1}{x-f_{1,1}}+ \frac{1}{x-f_{2,1}}& i=1, k = 1,\\
	\frac{1}{x-f_{1,2}}+ \frac{1}{x-f_{2,2}}& i=1, k = 2,\\
	0	& i=2,\\
	\end{cases} \label{eq:example_1_1}\\
	b_{j,l,k}(x) &= \omega_k(x)z^b_{j,l,k} 
	+ \omega_k(x) \times
	 \begin{cases}
		\frac{1}{x-f_{j,k}}& l=k,\\
		0	& else,\\
	\end{cases} 
	\label{eq:example_1_2}
\end{align}
\levi{for $i \in [2], j \in[2], l\in[2], k \in[2]$}  where $\omega_k(x) = (x-f_{1,k})(x-f_{2,k})$ and  $z^a_{i,l,k}$, $z^b_{j,l,k}$ are random noise terms that are independently and uniformly chosen from $\mathbb{F}_q$.
The master assigns each worker $g\in [N]$ a distinct element $x_g$ from $\mathbb{F}_q\setminus \{f_{1,1}, f_{1,2},f_{2,1}, f_{2,2}\}$. Thus, the query $\bfq_g, g\in [N]$ that the master sends to worker $g$ contains the evaluations of the encoding functions $\{a_{i,k}(x_g)\}_{i \in[2],k\in[2]}$ and $\{b_{j,l,k}(x_g)\}_{j \in[2],l\in[2],k\in[2]}$, the partitioning parameters, and the number of groups. Note that each encoding function contains a uniformly random variable. By Shamir's well-known secret sharing scheme \cite{shamir1979share}, each worker cannot gain any information about the coefficients in the encoding functions and, thus, cannot learn anything about $\mathcal{S}$.  Hence, the scheme is $T=1$ private.

After receiving $\bfq_g$, worker $g$ then encodes the matrices using
\begin{align} \label{eq:encoding}
	\hA_{k} = \sum_{i=1}^{2}\bfA_{i} a_{i,k}(x_g) ,
	\hB_{k} = \sum_{j=1}^{2}\sum_{l=1}^{2}\bfB_{j,l} b_{j,l,k}(x_g)
\end{align}
for $k\in[2]$. Now, the worker will calculate $\bfC(x_g) = \hA_{1}\hB_{1} + \hA_{2}\hB_{2}$ where the terms can be simplified into the following form:
\begin{align}\label{eq:example_2}
	\bfC(x_g)  &= \frac{\bfA_{1}\bfB_{1,1}}{(x_g-f_{1,1})} + \frac{\bfA_{1}\bfB_{2,1}}{(x_g - f_{2,1})}  \notag \\
	&+ \frac{\bfA_{1}\bfB_{1,2}}{(x_g-f_{1,2})} + \frac{\bfA_{1}\bfB_{2,2}}{(x_g - f_{2,2})} + \mathbf{I}(x_g)
\end{align}
where $\mathbf{I}(x)$ is a polynomial matrix that contains all the polynomial terms in $\bfC(x_g)$. Note that the maximum degree of $\mathbf{I}(x)$ is $\max_{k \in [2]}(\deg(\omega_k(x))) +2T-2 =2+2*1-2= 2$ since the largest polynomial degree in $\hA_k$ and $\hB_k$ is $\deg(\omega_k(x))+T-1$ and $T-1$, respectively. We highlight the fact that all desired matrices are coefficients to unique rational terms in Eq. \eqref{eq:example_2}. We achieved this by encoding each term in a desired matrix product with a unique root in the denominator so that when $\bfC(x)$ is computed, the desired matrix product remains the only term with the unique root in the denominator. One can think of $\omega(x)$ as a filter where all desired terms are kept with the rational terms and all other terms are aligned into polynomial terms. 

By Lemma \ref{lemma:cauchy_vandermonde}, we can interpolate Eq. \eqref{eq:example_2} from $7$ worker outputs since the polynomial terms have $3$ coefficients and the rational terms have $4$ coefficients.  Thus, the recovery threshold is $7$. Now, since $C(x) \in \mathbb{F}^{\alpha\times\frac{\alpha}{2}}_q$, the NDC is $\frac{7}{4}$. Additionally, we see that to calculate $C(x)$ the worker had to encode the matrices with complexity $\mathcal{O}\left(L_A\alpha^2 + L_B*2*\frac{\alpha^2}{2}\right) = \mathcal{O}\left(6\alpha^2\right)$ and then multiply and add the results with complexity $\mathcal{O}\left(\alpha^3\right)$. Since we assume that $\alpha$ is very large, the NCC is $\mathcal{O}\left(\alpha^3\times\frac{1}{|\calS|\alpha^3}\right) = \mathcal{O}\left(\frac{1}{2}\right)$.

\section{Main Result}\label{sec:main_res}

We now present the main result of this paper. 

\begin{theorem}\label{theorem:achievability}
	Assume a distributed system with $N$ workers, a computation list $\mathcal{S}$, and $T$ colluding workers. For any positive integers $m,n,r$ such that $m|\alpha$, $n|\alpha$, $r|mn$, and $|\mathbb{F}_q| \geq |\calS|mn+N$, there exists a privacy preserving scheme for up to $T$ colluding workers that achieves the following system metrics:
	\begin{align}
		&\text{Recovery Threshold: }  R = (\frac{r+1}{r})|\mathcal{S}|mn  + 2T-1 \\
		&\text{NDC: }  D = \frac{R}{|\mathcal{S}|mn} = \frac{r+1}{r} + \frac{2T-1}{|\mathcal{S}|mn}\\
		&\text{NCC: }  C = \mathcal{O}\left(\frac{r}{|\mathcal{S}|mn}\right)
	\end{align}
	\lev{Additionally, $r$ is the number of groups and provides no information about $\calS$.}
\end{theorem}
\begin{remark}\label{remark:example}
	\lev{
	Looking back at the example, we see that it corresponds to the case when $|\mathcal{S}| = 2$, $n=r=2$, $m=1$, and $T=1$ which results in $R = 7$, $D = \frac{7}{4}$, and $C = \frac{1}{2}$. Additionally, we could have also chosen $r=1$ which would result in $R = 9$, $D = \frac{9}{4}$, and $C=\frac{1}{4}$. We see that by allowing matrix partitioning and grouping we can achieve a wide variety of system parameters without breaking privacy.}
\end{remark}
\begin{remark}
	We note that our construction has a factor of $mn$ in the recovery threshold due to the simple partitioning method we use. We use this partitioning method to simplify the presentation of our coding scheme.  More complex partitioning and encoding methods using bilinear-complexity (see \cite{yu2020entangled,zhu2022systematic}) can be supported by CSA codes as demonstrated in \cite{tauz2022variable} which can allow further flexibility in computation and communication.  
\end{remark}

\begin{figure}
	\centering
	\begin{subfigure}{0.49\linewidth}
			\includegraphics[page=4, width=\linewidth]{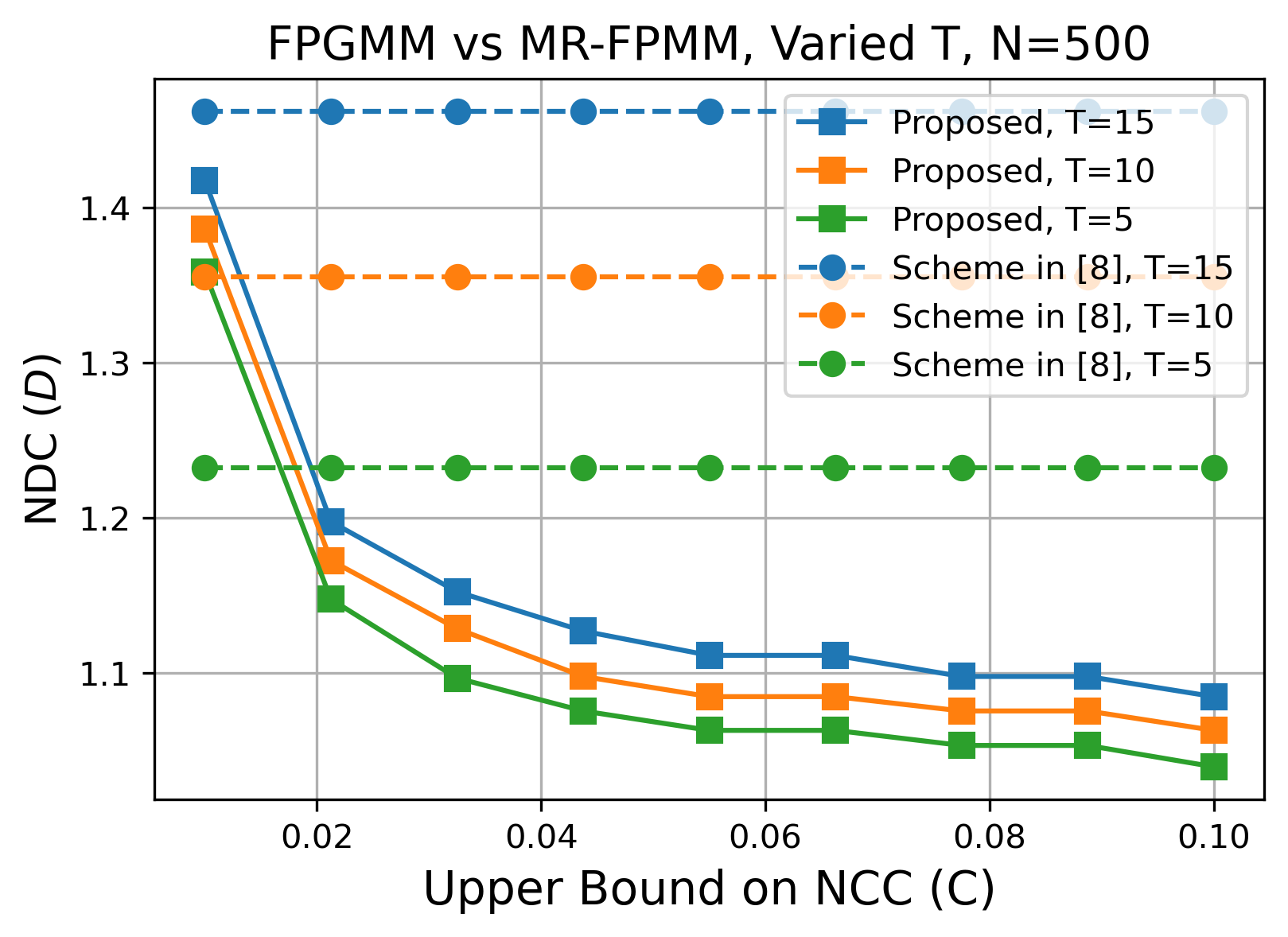}
			\caption{}\label{fig:c_v_d_500}
		\end{subfigure}
	\begin{subfigure}{0.49\linewidth}
			\includegraphics[page=4, width=\linewidth]{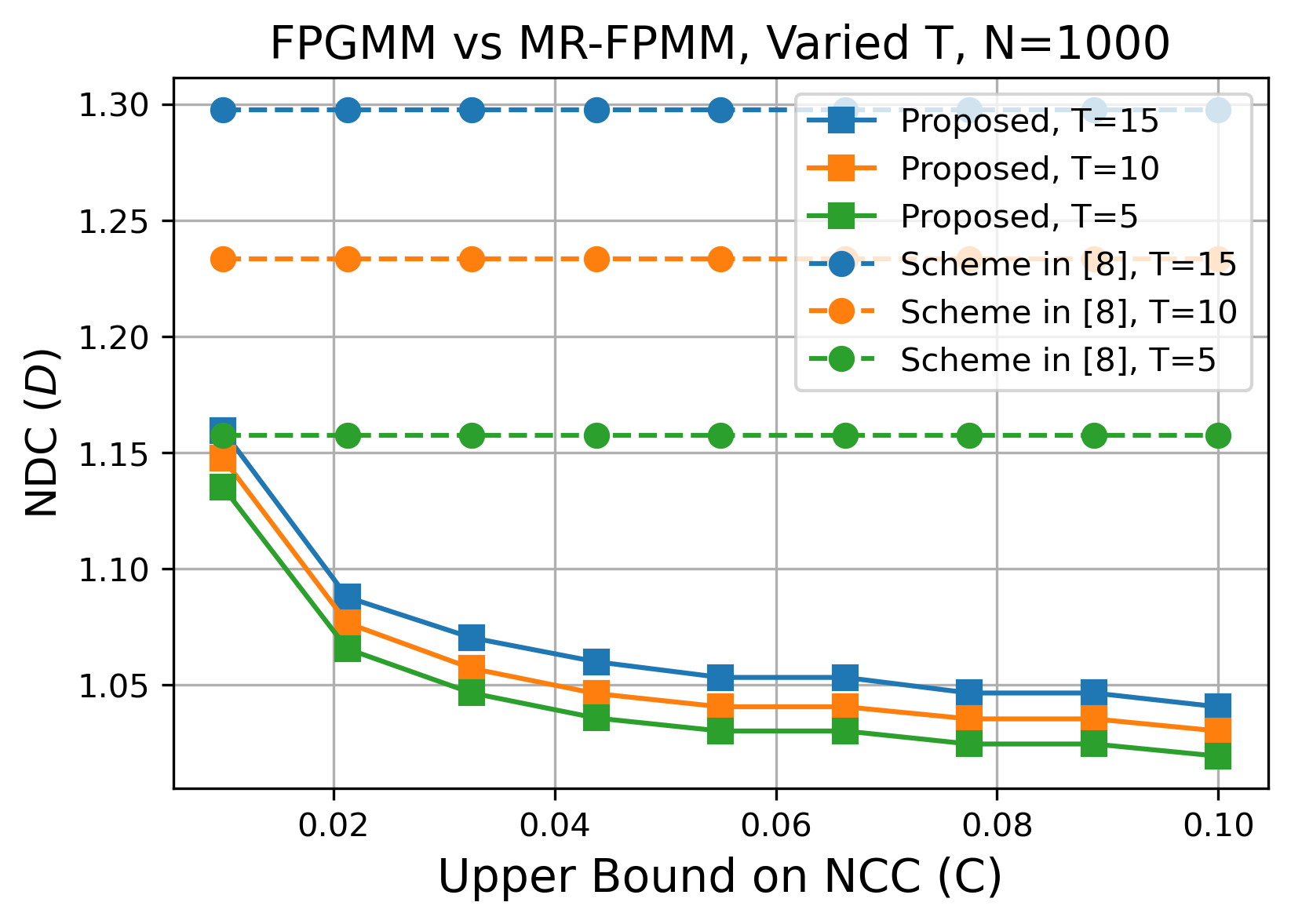}
			\caption{}\label{fig:c_v_d_1000}
		\end{subfigure}
	\caption{\levj{Comparison of FPGMM and MR-FPMM.  $|\mathcal{S}|$ is set to $5$. (a) Number of workers is upper bounded by 500. (b) Number of workers is upper bounded by 1000.}  }\label{fig:c_v_d}
	\vspace{-0.5cm}
\end{figure}

\begin{remark}
\levj{
For the non-straggler scenario, we now compare the trade-off of NCC and NDC of our scheme in comparison to a multi-round FPMM (MR-FPMM) scheme where in each round only one matrix product is computed. \lev{Note that a MR-FPMM scheme does not naturally preserve the privacy of the cardinality of $|\calS|$ but we shall ignore this for now to demonstrate how our scheme also provides benefits for the NCC and NDC. }  To the best of our knowledge, the best explicit scheme for FPMM is presented in  \cite{zhu2022systematic}\footnote{\lev{\cite{zhu2022systematic} also provides an implicit construction using bilinear complexity and Lagrange encoding. Due to space constraints, we focus on comparing our scheme with the explicit construction.}}. Given any three positive integers $(m,n,p)$ \lev{such that each divides $\alpha$}, the scheme in \cite{zhu2022systematic} has a recovery threshold of $ \widetilde{R} = \min((m+1)(np+T)-1, (n+1)(mp+T)-1, 2mnp+2T-1)$ where each worker uploads $\frac{\alpha^2}{mn}$ symbols to the master and performs $\mathcal{O}(\frac{\alpha^3}{mnp})$ computations at each worker for one round of FPMM. Thus, the NCC of this scheme is $\mathcal{O}(|\mathcal{S}|\frac{\frac{\alpha^3}{mnp}}{|\mathcal{S}|\alpha^3}) = \mathcal{O}(\frac{1}{mnp})$ and the NDC is $|\mathcal{S}|\frac{\frac{\widetilde{R}\alpha^2}{mn}}{|\mathcal{S}|\alpha^2} = \frac{\widetilde{R}}{mn}$ since the system has to operate $|\mathcal{S}|$ many times. 

Fig. \ref{fig:c_v_d} compares the trade-off between the NDC and NCC of our scheme versus the MR-FPMM scheme of \cite{zhu2022systematic}. For the experiment in Fig. \ref{fig:c_v_d}, we fix an upper bound value for NCC and then optimize the parameters $(m,n,r)$ of our scheme and $(m,n,p)$ of \cite{zhu2022systematic} in order to minimize the NDC while the NCC does not violate the upper bound  (i.e. any (x,y) coordinate in Fig. \ref{fig:c_v_d} indicates that this scheme has NDC of $y$ with an NCC of at most $x$). Additionally, we upper bound the number of workers so that the recovery threshold does not get arbitrarily large in order to reduce NDC. We see a striking difference between the two schemes since our proposed scheme shows a trend of decreasing NDC when NCC is increased and vice-versa while the multi-round scheme has a fixed NDC. This trend can be verified by looking at the NCC and NDC of the MR-FPMM scheme and seeing that the NCC and NDC are generally proportional to each other (after optimizing $(m,n,p)$)  and, thus, the NDC cannot improve with more computations for the MR-FPMM scheme. We also observe that our proposed scheme outperforms the multi-round scheme at almost all NCC values and offers more trade-off between the NCC and NDC. Additionally, we see that our scheme performs even better when the system has more workers, which can be see by comparing Fig.\ref{fig:c_v_d_500} and Fig.\ref{fig:c_v_d_1000}. Hence, our proposed scheme provides benefits for the performance metrics even when the cardinality of $|\calS|$ does not need to be protected. We close this remark with a reminder to the reader that this comparison was made to showcase the complexity benefits of our scheme and that MR-FPMM does not solve the privacy issues of FPGMM.} 
\end{remark}

\vspace{-0.2cm}
\section{Achievable Scheme: Proof of Theorem \ref{theorem:achievability}} \label{sec:scheme}
\vspace{-0.15cm}
\levj{
We now present our scheme for Theorem \ref{theorem:achievability} that achieves the stated system parameters. Assume that parameters $m,n,r$ are chosen according to Theorem \ref{theorem:achievability}.
}
\subsection{General Scheme}
\levj{
\subsubsection{Encoding Phase}
First, each query $\bfq_{g}, g\in[N]$ will tell each worker to equally partition each matrix in  $\bfA_{[L_A]}$ into $m$ sub-matrices in a row-wise manner, and the matrices in $\bfB_{[L_B]}$ into $n$ sub-matrices in a column-wise manner, i.e., $\bfA_i = \begin{bmatrix}
	\bfA^T_{i,1} &\cdots& \bfA^T_{i,m}
\end{bmatrix}^T \forall i \in [L_A]$ and $	\bfB_{j} = \begin{bmatrix}
\bfB_{j,1} &\cdots& \bfB_{j,n}
\end{bmatrix}, \forall j \in [L_B]$.
We can express all matrix products in $\ds$ as $\bfA_{i}\bfB_{j} = \{\bfA_{i,a}\bfB_{j,b}\}_{a \in [m],b\in[n]}$ for $(i,j) \in \calS$.
Thus, a sufficient condition to decode $\ds$ is to retrieve all matrix products $\dmn \overset{\triangle}{=}\{\bfA_{i,q}\bfB_{j,s}, (i,j) \in  \mathcal{S}, (q,s) \in  [m]\times [n]\}$. Note that this problem becomes another case of FPGMM with the key distinction that the workers know that $\dmn$ has at least $mn$ matrices, which does not reveal anything about $\mathcal{S}$. 
Thus, we can define a new FPGMM problem by re-indexing the sub-matrices into $\tA_i = \bfA_{\floor{\frac{i-1}{m}}+1,(i-1\mod m)+1}$ for $i \in [mL_A]$ and $\tB_j = \bfB_{\floor{\frac{j-1}{n}}+1,(j-1 \mod n)+1}$ for $j\in [nL_B]$ and specifying $\widetilde{\calS} = \{m(i-1)+q,n(j-1)+s: (i,j) \in  \mathcal{S}, (q,s) \in  [m]\times [n]\}$. Thus, we only consider $\tS$ for the rest of the paper.

Next, the master creates encoding functions to send to each worker. The master partitions $\tS$ into $r$ equal, non-overlapping partitions of size $\delta = \frac{|\mathcal{S}|mn}{r}$ denoted by $\cq_1,\dots,\cq_r$ where we assume $r|mn$ and, thus, $\delta$ is an integer.
For notational convenience, let $\mathcal{A}^k_i \subseteq \cq_k$ be the subset where the left index is $i$ and $\mathcal{B}_j^k \subseteq \cq_k$ be the subset where the right  index is $j$ for $i \in[mL_A]$,$j\in [nL_B],k \in [r]$. 
 Note that $\calA^k_{i} \cap \calB^k_{j} = \{(i,j)\}$ if $(i,j) \in \cq_k$ otherwise $\calA^k_{i} \cap \calB^k_{j}  = \emptyset$.
}

\lev{
The master then associates for each  $(i,j) \in \tS$ a distinct element $f_{i,j}$ from $\mathbb{F}_q$. We define the noise polynomials $a^n_{i,k}(x)$ for $i\in [mL_A],k\in[r]$ and $b^n_{j,k}(x)$ for $j \in [nL_B],k\in[r]$ as 
\begin{align}
	a^n_{i,k}(x) = \sum_{t \in [T]}z^a_{i,k,t}x^{t-1}, b^n_{j,k}(x)= \sum_{t\in [T]}z^b_{j,k,t}x^{t-1}, 
\end{align}
where $z^a_{i,k,t},z^b_{j,k,t}$ for $t \in [T],k \in [r], i \in [mL_A],j\in[nL_B]$ are random noise terms chosen independently and uniformly from $\mathbb{F}_q$.
Define the polynomial $\omega_k(x)$ as 
\begin{align}
	\omega_k(x) = \prod_{(i,j) \in \cq_k}(x-f_{i,j}) \label{eq:omega_def}
\end{align}
for $k \in[r]$. Note that $\omega_k(x)$ is a polynomial of degree $|\cq_k| = \delta$.
Next, the master defines the encoding functions $a_{i,k}(x)$ for $i\in [mL_A],k\in[r]$ and $b_{j,k}(x)$ for $j \in [nL_B],k\in[r]$ as
\begin{align}
	a_{i,k}(x) &= \omega_k(x) \left(\sum_{(q,s) \in \calA^k_{i}} \frac{1}{(x - f_{q,s})} + a^n_{i,k}(x)\right) \\
	b_{j,k}(x) &= \sum_{(q,s) \in \calB^k_j} \frac{1}{(x - f_{q,s})} + b^n_{j,k}(x).
\end{align}
}
 Now, note that 
\begin{subequations}\label{eq:multi}
	\begin{align}
		&a_{i,k}(x)b_{j,k}(x) = \sum_{(q,s) \in \calA^k_{i}}\sum_{(a,b) \in \calB^k_j} \frac{\omega_k(x)}{(x - f_{q,s})(x - f_{a,b})}  \label{eq:multi_a}\\
		&+\sum_{(q,s) \in \calA^k_{i}} \frac{\omega_k(x)b^n_{j,k}(x)}{(x - f_{q,s})} + \sum_{(q,s) \in \calB^k_j} \frac{\omega_k(x)a^n_{i,k}(x)}{(x - f_{q,s})} \label{eq:multi_b} \\
		&+ \omega_k(x)a^n_{i,k}(x)b^n_{j,k}(x). \label{eq:multi_c} 
	\end{align} 
\end{subequations}
Observe that since $\omega_k(x)$ has a zero root for all $x \in \{f_{i,j}\}_{(i,j) \in \cq_k}$, Eqs. \eqref{eq:multi_b} and \eqref{eq:multi_c} together are polynomials where the maximum degree of $\delta+2T-2$ comes from Eq. \eqref{eq:multi_c}. Now, we can re-formulate Eq. \eqref{eq:multi_a} as
\begin{subequations}
	\begin{align}
		&\eqref{eq:multi_a} = \sum_{(q,s) \in \calA^k_{i} \cap \calB^k_j} \frac{\prod_{(a,b) \in \cq_k\setminus (q,s)}(x-f_{a,b})}{(x - f_{q,s})} \label{eq:frac_a}\\
		&+\sum_{(q,s) \in \calA^k_{i}}\sum_{(a,b) \in \calB^k_j}\prod_{(c,d) \in \cq_k \setminus \{(q,s),(a,b)\}} (x-f_{c,d}). \label{eq:frac_b}
	\end{align}
\end{subequations}
Again, we see that Eq. \eqref{eq:frac_b} is a polynomial of maximum degree $\delta-2$ which is a smaller degree than the other polynomials in Eq. \eqref{eq:multi_c}. Now, consider Eq. \eqref{eq:frac_a}. If $(i,j) \notin \cq_k$, then Eq. \eqref{eq:frac_a} does not exist by the properties of $\calA^k_{i},\calB^k_{j}$. If $(i,j) \in \cq_k$, Eq. \eqref{eq:frac_a} can be expanded into a weighted sum of $(x-f_{i,j})^{-1},1,x,\dots,x^{\delta-2}$ by partial fraction decomposition, i.e. 
\begin{align}
	\frac{\prod_{(a,b) \in \cq_k\setminus (i,j)}(x-f_{a,b})}{(x - f_{i,j})} &= \frac{\gamma^{i,j,k}_{-1}}{(x-f_{i,j})} + \sum_{c=0}^{\delta-2}\gamma^{i,j,k}_c x^{c},
\end{align}
for constants $\{\gamma^{i,j,k}_{c}\}_{c=-1}^{\delta-2}$. Note that $\gamma^{i,j,k}_{-1} \neq 0$ otherwise Eq. \eqref{eq:frac_a} can be reduced further. Thus, we have
\begin{align}\label{eq:res_from_wab}
	a_{i,k}(x)b_{j,k}(x) = \beta_{i,j,k}(x) 
	+ \begin{cases}
		\frac{\gamma^{i,j,k}_{-1}}{(x-f_{i,j})} & (i,j) \in \cq_k, \\
		0 &  \text{ otherwise,}
	\end{cases}
\end{align}
where $\beta_{i,j,k}(x)$ is the sum of all polynomial terms in Eq. \eqref{eq:multi} and has a maximum degree of $\delta+2T-2$.

The master uses these encoding functions to create the queries $\bfq_i$. The master associates each worker $g \in [N]$ with a distinct element $x_g$ from $\mathbb{F}_q \setminus \{f_{i,j}, (i,j) \in \tS\}$. The query $\bfq_g,g\in[N]$ contains  $\{\{a_{i,k}(x_g)\}_{i=1}^{mL_A}\}_{k=1}^{r}$,  $\{\{b_{j,k}(x_g)\}_{j=1}^{nL_B}\}_{k=1}^{r}$, the number of groups $r$, and the partitioning parameters $m,n$. 

\subsubsection{Query and Computation}
Now, consider a worker $g$ for $g\in [N]$. Using $\bfq_g$, the worker creates the encoded matrices
\begin{align} \label{eq:encoding}
	\hA_{k} = \sum_{i=1}^{mL_A}\tA_{i} a_{i,k}(x_g) ,
	\hB_{k} = \sum_{j=1}^{nL_B}\tB_{j} b_{j,k}(x_g)
\end{align}
for $k\in[r]$. Note that $\hA_k$ and $\hB_k$ are encoded for the group $\cq_k$.  Now, the worker computes 
\begin{align} \label{eq:prod_sum}
	\sum_{k=1}^{r}\hA_{k}\hB_{k} &= \sum_{k=1}^{r}\sum_{i=1}^{mL_A}\sum_{j=1}^{nL_B} \tA_{i}\tB_{j} a_{i,k}(x_g)b_{j,k}(x_g) 
\end{align}
\begin{align}
	&=\sum_{k=1}^{r} \left( \sum_{(i,j) \in \cq_k}\frac{\gamma^{i,j,k}_{-1}\tA_{i}\tB_{j}}{(x_g-f_{i,j})} +  \sum_{i=1}^{mL_A}\sum_{j=1}^{nL_B}\tA_{i}\tB_{j}\beta_{i,j,k}(x_g) \right ) \\
	&= \sum_{(i,j) \in \tS } \frac{\gamma^{i,j,k}_{-1}\tA_{i}\tB_{j}}{(x_g-f_{i,j})} + \mathbf{I}(x_g) \overset{\triangle}{=} \bfU_g \label{eq:final_res}
\end{align}
where $\mathbf{I}(x)$ is a polynomial matrix of maximum degree $\delta+2T-2$ by Eq.\eqref{eq:res_from_wab}.
\subsubsection{Reconstruction}
Note that $\bfU_g$ is an evaluation of the function in Eq. \eqref{eq:final_res} at $x_g$ which is a sum of $|\tS| = |\mathcal{S}|mn$ rational terms and a polynomial of degree $\delta+2T-2$.  By Lemma \ref{lemma:cauchy_vandermonde},  Eq. \eqref{eq:final_res} can be interpolated from $|\mathcal{S}|mn+\delta + 2T-2 + 1 = (\frac{r+1}{r})|\mathcal{S}|mn  + 2T-1 = R$ worker outputs. Thus, we can retrieve  $\{\gamma^{i,j,k}_{-1}\tA_{i}\tB_{j}\}_{(i,j) \in \tS}$ from which we can easily extract $\{\tA_{i}\tB_{j}\}_{(i,j) \in \tS}$ since $\gamma^{i,j,k}_{-1} \neq 0$. Hence, $\ds$ is recoverable from any $R$ outputs.

%

\subsection{Privacy}
The full proof of privacy is provided in the Appendix. The main idea is that we are utilizing Shamir's secret sharing scheme \cite{shamir1979share,zhu2022private} to encode the components of the queries that will be used to  encode the data. By utilizing a $T-1$ degree random polynomial to encode the components, at least $T+1$ workers are required to extract any new information from the queries. \levj{We note that $\{f_{i,j}, (i,j) \in \tS\}$ is not provided to the workers otherwise they can infer the size of $\calS$.}
Thus, our scheme is privacy preserving up to $T$ colluding workers. 
\subsection{System Complexities and Decoding Complexity}
 Now, we discuss the NCC and NDC of the scheme. Due to partitioning, $\hA_{k}\in \mathbb{F}_q^{\frac{\alpha}{m} \times \alpha}$ and  $\hB_{k}\in \mathbb{F}_q^{\alpha \times \frac{\alpha}{n}}$ for $k \in [r]$. Thus, the complexity of encoding the matrices in Eq. \eqref{eq:encoding} is $\mathcal{O}(mL_A\frac{\alpha^2}{m} + nL_B\frac{\alpha^2}{n}) = \mathcal{O}((L_A+L_B)\alpha^2)$ and the complexity of calculating Eq. \eqref{eq:prod_sum} is $\mathcal{O}(\frac{\alpha^3r}{mn}+r\alpha^2)$. Since $\alpha \gg L_A,L_B$, the dominant complexity is the matrix multiplications with complexity $\mathcal{O}(\frac{\alpha^3r}{mn})$ and normalizing this term by  $|\mathcal{S}|\alpha^3$ results in the stated NCC. We can easily calculate the NDC by noting that $\bfU_{i} \in\mathbb{F}_q^{\frac{\alpha}{m} \times \frac{\alpha}{n}}$ and, thus, each worker outputs $\frac{\alpha^2}{mn}$ symbols. Hence, the worker retrieves $\frac{R\alpha^2}{mn}$ symbols that when normalized by $|\mathcal{S}|\alpha^2$ results in the stated NDC.
 
\lev{For completeness, we now discuss the decoding complexity at the master.} To extract the desired coefficients, the master has to interpolate Eq. \eqref{eq:final_res} for each entry in the matrix function of which there are $\frac{\alpha^2}{mn}$ entries. It is known that the complexity of interpolating an equation of the form Eq. \eqref{eq:cauchy_van_function} with $N$ rational terms and $K$ polynomial terms is $\mathcal{O}((N+K)\log^2(N+K)\log\log(N+K))$ \cite{jia2021cross,finck1993inversion}. Thus, the computational complexity of decoding is $\mathcal{O}(\frac{\alpha^2}{mn}R\log^2(R)\log\log(R))$. We remark that despite not strictly utilizing polynomials, we can achieve a decoding complexity comparable to polynomial interpolation which is a popular decoding method in coded computation literature \cite{aliasgari2020private,yu2020entangled,zhu2022systematic}.

\vspace{-0.1cm}

\section{Conclusion}\label{sec:conclusion}

In this paper, we present the fully private grouped matrix multiplication problem as a generalization of fully private matrix multiplication. We provide an achievable scheme for FPGMM that allows for flexibility between communication and computation cost while guaranteeing privacy. Additionally, we demonstrate that our scheme can outperform multi-round FPMM schemes for many operational points. Possible future work is improving the system metrics further by utilizing the natural redundancy in $\mathcal{S}$ since the same matrix may be used in multiple matrix products, as recently utilized in \cite{tauz2022variable}. 

\vspace{-0.5em}


\newpage
\bibliographystyle{ieeetr}
\bibliography{references}

\newpage
\appendix 
	
\subsection{Proof of Privacy}

We shall now prove that the proposed scheme satisfies the privacy requirement of Eq. \eqref{eq:privacy_req} for up to $T$ colluding workers. Note that if the scheme is privacy preserving against exactly $T$ colluding workers, then it is also privacy preserving against any $<T$ colluding workers since removing random variables cannot increase the mutual information. Thus, without loss of generality, let $\calT \in [N],|\calT| = T$.  From Eq. \eqref{eq:privacy_req}, we have
\begin{align}
	I(\mathcal{S}; \mathbf{q}_{\mathcal{T}}, \bfA_{[L_A]}, \bfB_{[L_B]}) &= I(\mathcal{S}; \bfA_{[L_A]}, \bfB_{[L_B]}) \notag \\
	&+ I(\mathcal{S}; \bfq_{\calT} |  \bfA_{[L_A]}, \bfB_{[L_B]}) \label{eq:private_proof_1}\\
	&= I(\mathcal{S}; \bfq_{\calT} ) \label{eq:private_proof_2}
\end{align}
where \eqref{eq:private_proof_1} comes from the chain rule of mutual information and \eqref{eq:private_proof_2} comes from the assumption that the data in $\bfA_{[L_A]}, \bfB_{[L_B]}$ is independent of the choice of $\mathcal{S}$ and that $\bfq_{\calT}$ is constructed without any information about $\bfA_{[L_A]}, \bfB_{[L_B]}$.  Now, we remind the reader that each $\bfq_{g}, g\in[N]$ contains the following components: 1) the parameters $m$ and $n$ to specify the matrix partitioning; 2) the parameter $r$ which specifies the number of groups; 3) The evaluations of the encoding functions $\{\{a_{i,k}(x_g)\}_{i=1}^{mL_A}\}_{k=1}^{r}$ and $\{\{b_{j,k}(x_g)\}_{j=1}^{nL_B}\}_{k=1}^{r}$. Component 1) is clearly independent of $\mathcal{S}$ since $m$, $n$ are system parameters. Additionally, component 2) is independent of $\calS$ because  the number of groups $r$ only depends on $mn$ and, thus, the workers cannot infer the size of $\calS$ since the reduced partitioned FPGMM problem contains at least $mn$ matrices in $\tS$. As such, we have
\begin{align}
	\eqref{eq:private_proof_2} &= I(\mathcal{S}; \underbrace{\{\{\{a_{i,k}(x_g)\}_{i=1}^{mL_A}, \{b_{j,k}(x_g)\}_{j=1}^{nL_B}\}^r_{k=1}\}_{g\in\calT}}_{\bfOMG_{\calT}}) \\
	&=  I(\mathcal{S}; \bfOMG_{\calT})  \\
	&= H(\bfOMG_{\calT}) - H(\bfOMG_{\calT}|\calS) \label{eq:private_inq 1}\\
	&\leq  H(\bfOMG_{\calT}) - H(\bfOMG_{\calT}|\calS, \{f_{i,j}\}_{(i,j)\in \tS}, \{x_g\}_{g\in \calT}) \label{eq:private_inq 2} \\
	&= H(\bfOMG_{\calT}) - H(\bfOMG_{\calT}|\calS, \{f_{i,j}\}_{(i,j)\in \tS},\{x_g\}_{g\in \calT}) \notag \\
	& + \underbrace{H(\bfOMG_{\calT}|\calS, \{f_{i,j}\}_{(i,j)\in \tS}, \{x_g\}_{g\in \calT}, \bfz)}_{=0} \label{eq:private_inq 3}\\
	&=  H(\bfOMG_{\calT}) - I(\bfz ; \bfOMG_{\calT} | \calS, \{f_{i,j}\}_{(i,j)\in \tS} ,\{x_g\}_{g\in \calT}) \label{eq:private_inq 4} \\
	&=  H(\bfOMG_{\calT}) - H(\bfz|\calS, \{f_{i,j}\}_{(i,j)\in \tS} ,\{x_g\}_{g\in \calT} ) \notag \\
	&+ H(\bfz|\calS, \{f_{i,j}\}_{(i,j)\in \tS} ,\{x_g\}_{g\in \calT},\bfOMG_{\calT} ) \label{eq:private_inq 5}
\end{align}
where $\bfz = \{\{\{z^a_{i,k,t}\}_{i=1}^{mL_A}, \{z^b_{j,k,t}\}_{j=1}^{nL_B}\}_{k=1}^r\}^T_{t=1}$,  \eqref{eq:private_inq 2} comes from the fact that conditioning reduces entropy,  and  \eqref{eq:private_inq 3} comes from $\bfOMG_{\calT}$ being a deterministic function of $\calS, \{f_{i,j}\}_{(i,j)\in \tS}, \{x_g\}_{g\in \calT}, \bfz)$. Recall that $\{\{a^n_{i,k}(x)\}_{i=1}^{mL_A}, \{b^n_{j,k}(x)\}_{j=1}^{L_B}\}^r_{k=1}$ are the random polynomials with coefficients in $\bfz$. We note that 
\begin{align}
	&H(\bfz|\underbrace{\calS, \{f_{i,j}\}_{(i,j)\in \tS} ,\{x_g\}_{g\in \calT},\bfOMG_{\calT}}_{\mathbf{P}} ) \label{eq:h_cond_1}\\
	&= H(\bfz|\mathbf{P} , \{\{\{a^n_{i,k}(x_g)\}_{i=1}^{mL_A}, \{b^n_{j,k}(x_g)\}_{j=1}^{L_B}\}^r_{k=1}\}_{g\in\calT}) \label{eq:h_cond_2} \\
	&=H(\bfz|\{\{\{a^n_{i,k}(x_g)\}_{i=1}^{mL_A}, \{b^n_{j,k}(x_g)\}_{j=1}^{L_B}\}^r_{k=1}\}_{g\in\calT}) \label{eq:h_cond_3} \\
	&= 0 \label{eq:h_cond_4}
\end{align}
where \eqref{eq:h_cond_2} comes from the fact that $\{\{\{a^n_{i,k}(x_g)\}_{i=1}^{mL_A}, \{b^n_{j,k}(x_g)\}_{j=1}^{L_B}\}^r_{k=1}\}_{g\in\calT}$ can be calculated by removing the rational terms from $\bfOMG_{\calT}$ using $\calS, \{f_{i,j}\}_{(i,j)\in \tS} ,$ and $\{x_g\}_{g\in \calT}$ leaving only the noise polynomials and \eqref{eq:h_cond_4} comes from the fact that $a^n_{i,k}(x),b^n_{j,k}(x)$ are polynomials in $x$ with maximum degree $T-1$ which can be interpolated from the $T$ evaluations to determine $\bfz$. Now, we have that 
\begin{align}
	\eqref{eq:private_inq 5} &=  H(\bfOMG_{\calT}) - H(\bfz|\calS, \{f_{i,j}\}_{(i,j)\in \tS} ,\{x_g\}_{g\in \calT} )  \\
	&= H(\bfOMG_{\calT}) - H(\bfz) \label{eq:private_small_eq_1} \\
	&\leq 0 \label{eq:private_small_eq_2} 
\end{align}
where \eqref{eq:private_small_eq_1} comes from the fact that the random variables in $\bfz$ are independent from $\mathcal{S},\{f_{i,j}\}_{(i,j)\in \tS} ,\{x_g\}_{g\in \calT}$ and \eqref{eq:private_small_eq_2} comes from $\bfOMG_{\calT}$ and $\bfz$ having the same number of symbols and the symbols in $\bfz$ are generated using the maximum entropy distribution implying that $H(\bfOMG_{\calT}) \leq H(\bfz)$. Thus, we have proven that $I(\mathcal{S}; \mathbf{q}_{\mathcal{T}}, \bfA_{[L_A]}, \bfB_{[L_B]})\leq 0$ which completes the proof since mutual information is non-negative. 
\end{document}